\documentclass[final]{IEEEtran}
\ifCLASSINFOpdf
\else
\fi

\usepackage{amsthm,amssymb,graphicx,multirow,amsmath,bm, color,amsfonts}%,ulem}
\usepackage[update,prepend]{epstopdf}
\usepackage{cite}
\usepackage{tabulary}
\usepackage{bbm} % for \mathbbm{1}
\usepackage{multirow}
\usepackage{comment}
\usepackage[caption=false]{subfig}
\usepackage[ruled,linesnumbered]{algorithm2e}
\SetKw{KwBy}{by}
\usepackage{cancel} 
\usepackage[normalem]{ulem}

\usepackage{booktabs}
\usepackage[table,xcdraw]{xcolor}

\usepackage{changepage}  % put in preamble

\makeatletter
\newcommand{\nosemic}{\renewcommand{\@endalgocfline}{\relax}}% Drop semi-colon ;
\newcommand{\dosemic}{\renewcommand{\@endalgocfline}{\algocf@endline}}% Reinstate semi-colon ;
\let\oldnl\nl% Store \nl in \oldnl
\newcommand{\nonl}{\renewcommand{\nl}{\let\nl\oldnl}}% Remove line number for one line
\usepackage{makecell}

\usepackage{optidef}

\def\nb0{{\mathbf{0}}}
\def\nb1{{\mathbf{1}}}

\def\figref#1{Fig.\,\ref{#1}}%

\allowdisplaybreaks % Allows breaking of eqnarray over multiple pages (avoids unnecessary blanks in the document before eqnarray)
\begin{document}

\pagenumbering{gobble}
\graphicspath{{./figures/}}
\title{
%	MAINTAINED: Autonomous Artificial Intelligence Agent for Wireless Network Deployment
Small Models, Big Impact: Tool-Augmented AI Agents for Wireless Network Planning
}
\author{
	Yongqiang Zhang, \IEEEmembership{Graduate Student Member,~IEEE,} 
	Mustafa A. Kishk, \IEEEmembership{Member,~IEEE,} and Mohamed-Slim~Alouini, \IEEEmembership{Fellow,~IEEE}
	\thanks{Yongqiang Zhang and Mohamed-Slim Alouini are with Computer, Electrical and Mathematical Science
		and Engineering Division, King Abdullah University of Science and Technology (KAUST), Thuwal 23955-6900, Kingdom of Saudi Arabia. Email: \{yongqiang.zhang.2, slim.alouini\}@kaust.edu.sa.}
	\thanks{Mustafa A. Kishk is with the Department of Electronic Engineering, Maynooth University, Maynooth, W23 F2H6, Ireland. Email: mustafa.kishk@mu.ie.}
}

\maketitle
\begin{abstract}
Large Language Models (LLMs) such as ChatGPT promise revolutionary capabilities for Sixth-Generation (6G) wireless networks  but their massive computational requirements and tendency to generate technically incorrect information create deployment barriers.
In this work, we introduce \textsc{MAINTAINED}: autono\underline{m}ous \underline{a}rtificial \underline{int}elligence \underline{a}gent for w\underline{i}reless \underline{ne}twork \underline{d}eployment. 
Instead of encoding domain knowledge within model parameters, our approach orchestrates specialized computational tools for geographic analysis, signal propagation modeling, and network optimization.
In a real-world case study,  MAINTAINED outperforms state-of-the-art LLMs including ChatGPT-4o,  Claude Sonnet 4,  and DeepSeek-R1 by up to 100-fold in verified performance metrics while requiring less computational resources.
This paradigm shift, moving from relying on parametric knowledge towards externalizing domain knowledge into verifiable computational tools, eliminates hallucination in technical specifications and enables edge-deployable Artificial Intelligence (AI) for wireless communications.
\end{abstract}
% Note that keywords are not normally used for peerreview papers.

\section{Introduction}\label{sec:intro}
The vision of 6G wireless networks promises ubiquitous intelligence where AI capabilities are seamlessly integrated throughout network infrastructure \cite{Bariah2024a}.
LLMs such as ChatGPT, Claude, and DeepSeek have demonstrated remarkable cognitive abilities in complex reasoning and decision-making tasks, making them attractive candidates for intelligent network management and optimization. 
However, deploying these models for real-world wireless network planning faces a critical challenge: the {\em hallucination} problem, where models generate plausible but factually incorrect technical information.
In wireless network deployment, precision is paramount as even minor errors can have cascading consequences. 
Incorrect antenna specifications, erroneous coverage calculations, or invalid frequency allocations can lead to service outages, regulatory violations, and costly network failures. 
Evidence indicates that even advanced LLMs hallucinate technical specifications in 5–15\% of domain-specific queries,  a rate unacceptable in mission-critical deployments such as wireless network planning,  where inaccurate calculations can result in inadequate coverage and costly infrastructure failures \cite{Ji2023}.

\begin{table*}[htbp]
	\centering
	\caption{Current LAM Approaches and Their Limitations for Wireless Network Planning}
	\label{tab:lam_limitations}
	\footnotesize
	\setlength{\tabcolsep}{3pt}
	\renewcommand{\arraystretch}{0.9}
	\begin{tabular}{|p{2.2cm}|p{6cm}|p{7cm}|}
		\hline
		\textbf{Approach} & \textbf{Description \& Method} & \textbf{Limitations for Wireless Communications Domain} \\ 
		\hline
		Fine-tuning & Training pre-trained models on domain-specific wireless datasets to improve performance & Requires massive computational resources and extensive datasets. Impractical for rapidly evolving 6G standards \\
		\hline
		RAG & Augmenting model responses by retrieving relevant information from external knowledge bases & Cannot perform real-time calculations or execute computational tasks necessary for network optimization \\
		\hline
		In-context learning & Providing technical examples within prompts to guide model behavior without parameter updates & Requires large context windows and powerful models while remaining prone to hallucination when generating responses \\
		\hline
	\end{tabular}
\end{table*}

Existing approaches for LLM-based wireless network planning fall into three main categories: fine-tuning methods, Retrieval-Augmented Generation (RAG) approaches, and in-context learning techniques, each with fundamental drawbacks as outlined in Table~\ref{tab:lam_limitations} \cite{Zou2025, Zhang2024c, Zhou2025}.
Beyond their individual limitations, all these traditional approaches share two critical weaknesses. 
One is their reliance on encoding domain knowledge within model parameters, which renders them prone to generating technically inaccurate information when the embedded knowledge is incomplete.
Another is their substantial computational demands, which are fundamentally misaligned with 6G’s emphasis on edge intelligence.

Recently, the agentic AI paradigm is gaining significant traction in wireless communications~\cite{Jiang2025}.
Prior studies have explored the application of AI agents for solving wireless network planning problems, such as radio map generation, network slicing, and indoor network design~\cite{Quan2025,Hou2025 ,Tong2025}.
However, these initial approaches still operate largely as ``co-pilots," relying heavily on explicit prompt engineering and a ``human-in-the-loop" operational model to guide their execution. 
This reliance on human guidance highlights a critical gap in achieving autonomous AI, which recent LLM breakthroughs are poised to address.
Starting from OpenAI's introduction of {\em function calling} capabilities in GPT-3.5-turbo-0613, LLMs have gained the ability to interact with external tools and application programming interfaces (APIs), executing computational tasks beyond text generation.
Complex problem-solving scenarios often require adaptive decision-making, where models must analyze intermediate results to determine optimal next steps rather than following predetermined sequences.
The Reasoning and Acting (ReAct) framework enables such adaptive behavior by interleaving analytical reasoning with action execution, allowing models to dynamically adjust their strategies based on real-time observations  \cite{Yao2023}.
In this work, we present MAINTAINED, an alternative paradigm for building a domain-specific AI agent for wireless networks that achieves reliable wireless network deployment by externalizing domain knowledge into verifiable computational tools, rather than relying solely on parametric knowledge.
In particular, by leveraging function calling to externalize domain knowledge into verifiable computational tools, our framework eliminates the need for any domain-specific continual training or fine-tuning. Furthermore, the ReAct framework empowers MAINTAINED to operate as an autonomous agent, capable of autonomously decomposing and executing complex, multi-step deployment tasks from a single high-level user objective, thus removing the requirement for expert human-in-the-loop guidance.
Our main contributions are summarized as follows:
\begin{itemize}
	\item We introduce MAINTAINED, an autonomous agent that eliminates hallucination in wireless network planning by replacing parametric knowledge with verifiable computational tools,  achieving reliable technical decision-making through ReAct reasoning and function calling without large-scale model training.
	\item We demonstrate that a compact  4 billion (4B) parameter model with tool orchestration outperforms ChatGPT,  Claude, and DeepSeek in wireless network deployment,  achieving 100-fold better verified performance while using less computational resources. 
	\item  We provide an open-source implementation that demonstrates the agentic integration of multimodal data sources,  physics-based propagation models, and mathematical optimization tools, enabling edge-deployable AI planning for wireless networks.
\end{itemize}

\section{Function Calling and ReAct Agents}
For the application of large AI models in the wireless communications domain, the fundamental challenge lies in harnessing LLM reasoning for wireless engineering while simultaneously ensuring computational precision and avoiding technical hallucination.
This challenge can be addressed by two complementary technologies:  (i) function calling for verifiable computation; and (ii) the ReAct framework for intelligent orchestration of multiple function-calling tools.
\subsection{Function Calling} \label{sec:fn_call}

Function calling fundamentally changes how LLMs interact with technical systems. Instead of generating text that describes what a calculation might produce, function calling directly executes the calculation and returns verified results. This shift from description to computation eliminates hallucination in technical domains.
Following OpenAI’s introduction of function calling capabilities in GPT-3.5-turbo-0613 in June 2023, the feature quickly gained adoption among major LLM providers, with Anthropic integrating function calling into its Claude 3 models in March 2024, Alibaba adding support in Qwen 2.5 in September 2024, and Google introducing it for its Gemini models via Vertex AI.
This widespread adoption reflects the critical importance of computational grounding for technical applications.

\begin{figure}[!t]
	\centering
	\includegraphics[width= 0.8\columnwidth]{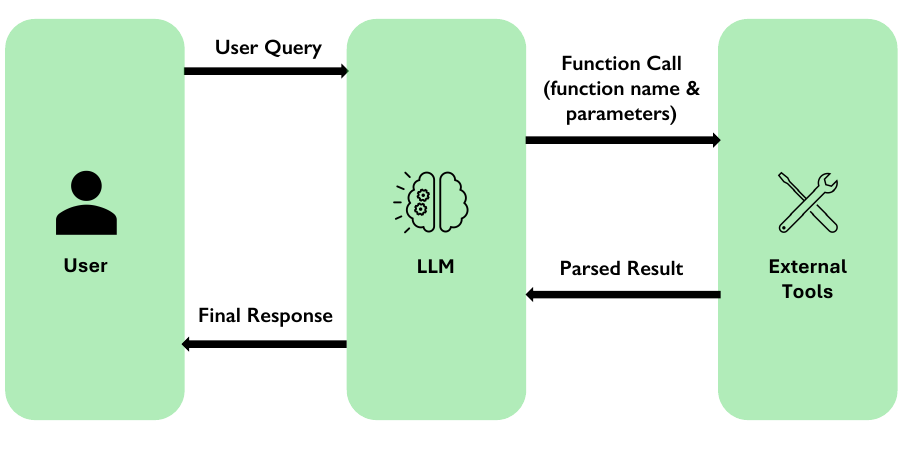}
	\caption{Illustration of function calling in LLM systems.}
	\label{fig:function_calling}
\end{figure}

\figref{fig:function_calling} illustrates the fundamental architectural shift enabled by function calling. Traditional LLMs generate responses by predicting the next token in a sequence, which can result in plausible but potentially inaccurate technical details. 
For example,  when asked \textit{``What is the received signal strength at 2 km from a 5 GHz base station with 20 dBm transmit power?''},  a conventional model might return an estimated answer like \textit{``-75 dBm''} based on typical propagation patterns learned during training.
In contrast, function calling enables the model to invoke an external computational tool (such as a path loss calculator) by inferring both the function name and the required input parameters from the user query. 
The final response is then based on the result of that actual computation, not on language modeling alone.
The technical implementation of function calling involves structured JavaScript Object Notation (JSON) schema definitions that specify precise input requirements, expected output formats, and functional descriptions for each available computational tool. 

It is important to note that function calling requires developers to implement the computational functions themselves. 
For wireless communications applications, this means writing functions for propagation calculations, antenna pattern analysis, spectrum allocation, and network optimization algorithms. 
This approach offers significant advantages over traditional fine-tuning methods: developers avoid preparing extensive labeled datasets and performing expensive model retraining.
Instead of encoding domain knowledge within model parameters through fine-tuning, function calling externalizes this knowledge into code modules that can be updated independently of the language model.
Moreover, this deterministic approach eliminates uncertainty in technical calculations while enabling models to access real-time data and specialized computational capabilities beyond their training scope. 
Specifically, function calling provides the following critical benefits that address fundamental limitations in technical AI applications.
\subsubsection{Zero Hallucination for Computational Results} 
Mathematical computations produce deterministic outputs based on verified algorithms. 
For example, path loss calculations use established ITU-R propagation models through custom functions, which provide guaranteed accuracy and eliminate the unpredictability inherent in token-based generation.

\subsubsection{Access to Up-to-Date Information}\label{sec:time}
Technical planning requires up-to-date information that extends beyond static training data. Function calling enables direct access to live regulatory databases, current equipment specifications, real-time spectrum monitoring, and dynamic network conditions. 

\subsubsection{Domain-Specific Tool Integration} 
Engineering applications require domain-specific tools for computational tasks and data modalities beyond language model capabilities.
Function calling enables seamless integration with existing libraries and custom functions designed for technical workflows. 
Integrating existing specialized tools enables efficient processing and generation of diverse data modalities that are impractical for text-based language model inference.

\subsubsection{Inference Cost Reduction}
Function calling significantly reduces inference costs by eliminating lengthy token sequences for complex calculations.
In cases where specialized computation is needed,  function calling shifts the burden from the language model to optimized external tools, which can reduce token usage and latency.

\subsection{ReAct Framework}

 \begin{figure}[!t]
	\centering
	\includegraphics[width=0.8\columnwidth]{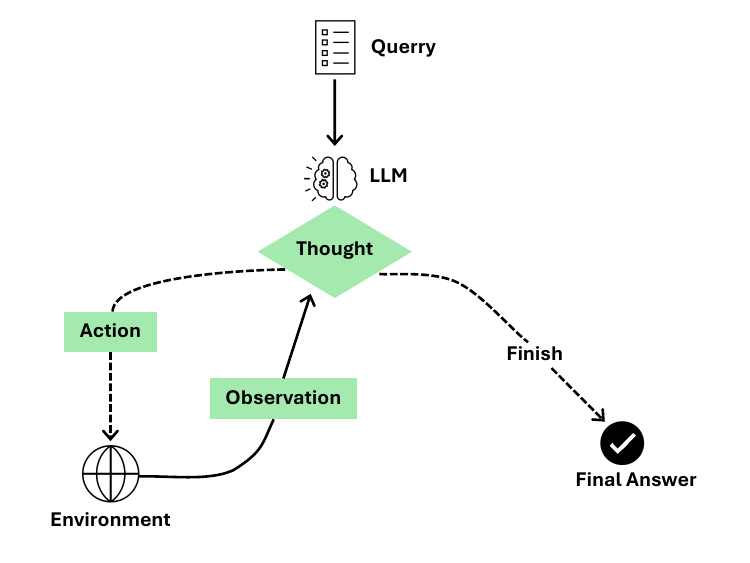}
	\caption{The ReAct cycle.}
	\label{fig:react_cycle}
\end{figure}

Traditional LLM approaches face fundamental limitations when tackling complex problems. 
Chain-of-thought reasoning alone, while enabling step-by-step thinking, suffers from hallucination and error propagation due to lack of external grounding. 
Conversely, action-based approaches that interact with environments lack systematic reasoning and planning capabilities. 
The ReAct framework addresses these limitations by synergistically combining reasoning and acting, where reasoning traces guide action selection while action outcomes ground and update the reasoning process.
For example, when asked \textit{``Deploy base stations for a 50,000-seat stadium,''} a conventional model without ReAct might return a straightforward response like \textit{``Install 200 small cells uniformly distributed across the venue''} based on typical deployment patterns learned during training.
In contrast,   ReAct agent would first reason \textit{``I need to analyze capacity requirements for different zones''}.
Then,  it acts by calculating density patterns and observes \textit{``Entrance areas show 2 times higher user density during event start/end times''}. 
This observation leads to further reasoning,  \textit{``Uniform distribution would create severe congestion at entrances''}, which prompts adaptive cell placement with higher density at entry points.
The final response emerges from this iterative reason-act-observe cycle,  rather than from blindly applying patterns memorized from the training data.

\figref{fig:react_cycle} illustrates the ReAct cycle's fundamental architecture. 
The framework transforms traditional LLM interaction by introducing structured decision-making loops that alternate between internal reasoning and external action execution. 
This architectural approach enables LLMs to maintain systematic problem decomposition while incorporating real-world feedback,  creating a bridge between pure language generation and grounded computational workflows.
As shown in the cycle, ReAct operates through three coordinated phases:
\begin{itemize}
	\item \textbf{Thought}: The agent generates reasoning traces that decompose the problem, analyze current progress, and determine optimal next steps. 
	These reasoning steps provide interpretability and help maintain context across multiple iterations. The thought phase leverages chain-of-thought capabilities while avoiding unconstrained reasoning that leads to hallucination.
	\item \textbf{Action}: Based on reasoning conclusions, the agent takes specific actions to gather information, execute computations, or interact with external systems.
	 These actions are grounded in the reasoning process and serve to advance toward the goal. 
	Actions can range from simple information retrieval to complex multi-step computational workflows.
	\item \textbf{Observation}: The agent incorporates feedback from actions into its understanding, updating its knowledge state and informing subsequent reasoning cycles. 
	This creates a feedback loop where actions inform reasoning and reasoning guides actions, enabling dynamic strategy adjustment based on intermediate results.
\end{itemize}

The technical implementation of ReAct involves prompt engineering techniques that structure language model behavior to follow this iterative pattern. Unlike function calling, which requires explicit schema definitions, ReAct can operate with more flexible prompting strategies that guide the model to naturally alternate between reasoning and acting phases. This flexibility enables adaptation to diverse problem domains without requiring extensive upfront specification of all possible actions.
However, effective ReAct implementation requires careful consideration of termination conditions, error handling, and reasoning quality assessment. The iterative nature can lead to infinite loops without proper safeguards, and the reasoning quality directly impacts overall performance. Additionally, the multi-step nature increases computational overhead compared to direct function calling approaches.
ReAct provides several important advantages that address fundamental limitations in complex problem-solving scenarios:

\subsubsection{Dynamic Adaptation and Error Recovery}
The iterative structure enables agents to recover from errors and adjust strategies based on intermediate feedback. 
When initial approaches fail or produce unexpected results, the reasoning phase can analyze the situation and modify the approach accordingly. 
This adaptability surpasses rigid procedural approaches that cannot respond to unforeseen circumstances.

\subsubsection{Transparent Decision-Making Process}
Unlike black-box approaches, ReAct maintains explicit reasoning traces that document decision rationale throughout the problem-solving process. 
This transparency facilitates debugging, verification, and understanding of agent behavior, which proves essential for technical applications requiring audit trails and quality assurance.

\subsubsection{Flexible Problem Decomposition}
ReAct naturally handles complex problems that require dynamic decomposition into subtasks. 
The reasoning phase can identify appropriate problem breakdowns based on current context, while the action phase can execute various computational strategies as needed. 
This flexibility enables handling of diverse problem types without predetermined algorithmic structures.

\subsubsection{Reduced Model Size Requirements}
By structuring problem-solving into explicit reasoning and acting phases,  ReAct enables models to tackle complex tasks more efficiently than attempting to solve them in a single forward pass. 
The iterative approach allows models to break down problems into manageable steps, potentially achieving better performance without requiring larger model sizes.
This structured decomposition can help smaller models systematically work through problems that might otherwise require more parameters to handle in an end-to-end fashion.

\section{The MAINTAINED Framework}\label{sec:3}
The MAINTAINED framework embodies a fundamental design philosophy that diverges from conventional large AI model approaches in wireless communications. 
Rather than attempting to encode domain expertise within model parameters through extensive training, we externalize specialized knowledge into computational tools while leveraging the LLM exclusively for orchestration and reasoning. 
This computation-over-memorization approach addresses critical limitations inherent in traditional LLM applications for wireless network planning, ensuring that all technical computations are performed by verified algorithms and external data sources while eliminating hallucination concerns.

\begin{figure*}[htbp]
	\centering
	\includegraphics[width=0.80\textwidth]{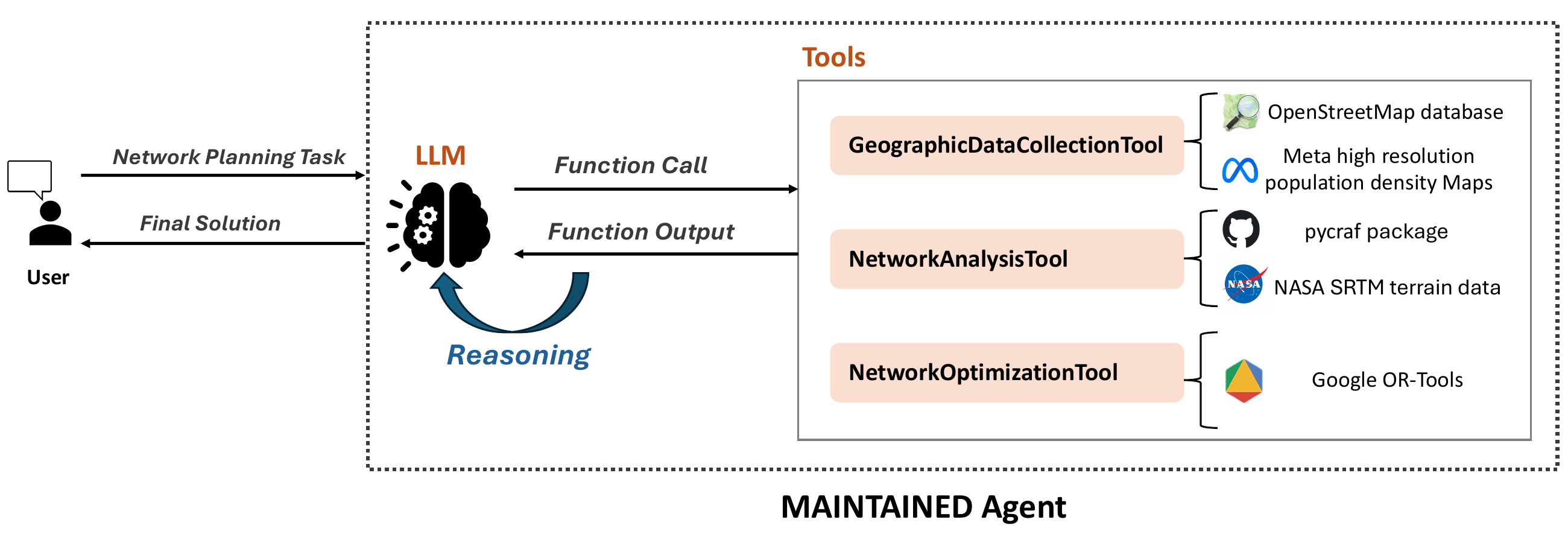}
	\caption{The architecture of our proposed MAINTAINED framework. }
	\label{fig:framework}
\end{figure*}

Fig.~\ref{fig:framework} illustrates the architecture of the MAINTAINED framework. 
Our proposed MAINTAINED framework consists of an LLM and a set of specialized tools.
The LLM interacts with these tools via function calling mechanisms: 
the \FuncSty{GeographicDataCollectionTool},  which leverages OpenStreetMap  databases and Meta’s high resolution population density maps; the \FuncSty{NetworkAnalysisTool},  which utilizes the \FuncSty{pycraf} package together with National Aeronautics and Space Administration
(NASA) Shuttle Radar Topography Mission (SRTM) terrain data for signal propagation modeling;
and the \FuncSty{NetworkOptimizationTool},  which employs Google OR-Tools for mathematical optimization \cite{Boeing2025, meta_hrsl_2023, winkel_pycraf_2018}.
This architecture embodies our computation-over-memorization philosophy, externalizing domain expertise into verified computational tools rather than encoding it in potentially unreliable model parameters. 
The LLM orchestrates these tools via function calls, dynamically decomposing complex deployment tasks and adapting strategies based on intermediate outcomes.
Within this framework, users can interact using natural language, while the agent ensures computational accuracy through verified algorithms and autonomously generates complete network deployment solutions without requiring human intervention.

\subsection{Task Specification Through Natural Language}\label{sec:task}
The MAINTAINED framework fundamentally transforms how network planners interact with deployment tools by enabling natural language task specification. Conventional wireless network planning tools typically require structured inputs, technical expertise, and manual configuration of system parameters such as transmit power, deployment region, data rate thresholds, and budget constraints. In contrast, our proposed agent removes these barriers by allowing users to describe network planning tasks directly in natural language.
The framework supports this interaction model by automatically parsing free-form user prompts into structured specifications required for downstream computation. This natural language interface processes diverse deployment requirements across four key dimensions:
\begin{itemize}
\item \textbf{Geographic Specifications}: The framework accepts latitude and longitude coordinates defining target deployment regions, enabling precise geographic boundary definition through conversational descriptions like ``the area between 21.0 and 21.5 degrees latitude."

\item \textbf{Service Requirements}: Users specify data rate thresholds, coverage objectives, and quality metrics through intuitive statements such as ``ensure minimum 2 Mbps per user," which are automatically translated into quantitative optimization constraints.

\item \textbf{Operational Constraints}: The system processes budget limitations and infrastructure specifications, incorporating constraints like ``HAPs cost around 1200 units and TBSs about 600 units" directly into the optimization framework.

\item \textbf{Performance Objectives}: Multi-objective optimization preferences are expressed through natural language, such as ``find the minimum-cost deployment plan", which the framework converts into appropriate optimization parameters.
\end{itemize} 

To illustrate the expressive power and accessibility of this approach, consider the following example of a complete network deployment task expressed as a single natural language prompt:
\begin{adjustwidth}{0.5cm}{0.5cm}
\small
\ttfamily
I'm working on a plan to set up a wireless network in a region of Saudi Arabia, specifically within the area between 21.0 and 21.5 degrees latitude, and 43.5 to 44.0 degrees longitude. The network will operate on the 5 GHz frequency band, using a 10 MHz channel bandwidth.

For the infrastructure, I'm considering a combination of High Altitude Platforms (HAPs) and Terrestrial Base Stations (TBSs). Each HAP is estimated to cost around 1200 units, and each TBS about 600 units. Both types of stations will transmit at 20 watts of power.

Could you help analyze this area for me? 
I’m particularly interested in understanding how much the total deployment would cost,  what kind of average data rates customers can expect, and how many HAPs and TBSs we’ll need to deploy. 
It would also be great to know where these should be placed, and whether all demand nodes can meet the minimum requirement of 2 Mbps. 
\end{adjustwidth}

This single natural language prompt demonstrates the MAINTAINED framework’s capabilities, acting as the sole required human input. 
The agent autonomously manages all subsequent tasks without further human intervention.
Throughout the following sections, we will show how the agent processes this single conversational prompt to generate a complete network deployment solution, coordinating multiple computational tools to transform high-level requirements into an optimized infrastructure plan. 
By tracking the agent's reasoning process and tool interactions for this real-world scenario, we illustrate how natural language interfaces can democratize advanced network optimization capabilities while maintaining computational rigor and accuracy. 

\subsection{Intelligent Tool Coordination}

Unlike traditional approaches that require predefined workflows, our framework enables dynamic tool coordination through ReAct reasoning. 
The agent does not follow a rigid pipeline.
Instead,  given an input task, it autonomously determines which tools to invoke and in what sequence based on the specific deployment scenario. 
This adaptive approach allows the agent to reason about task requirements and self-coordinate appropriate function calling traces step by step.

\subsubsection{\FuncSty{GeographicDataCollectionTool}}
This tool establishes the spatial foundation for network planning by identifying both demand locations and existing infrastructure within the specified deployment region through automated processing of authoritative sources. 
Given the geographic coordinates from the user prompt, it performs three critical functions. 
First,  it processes Meta's high resolution population density maps to identify where potential users are located, effectively translating population distributions into demand nodes for network planning. 
Second,  it queries OpenStreetMap databases to discover existing communication towers within the region,  providing crucial information about available infrastructure that can be leveraged. 
Third,  it generates optimized candidate location grids for both HAPs and TBSs based on the identified demand patterns. 
The tool produces visualizations of the deployment region alongside structured spatial information outputs.

\subsubsection{\FuncSty{NetworkAnalysisTool}}
This tool performs physics-based signal propagation analysis to determine the feasibility and quality of potential network links. 
Using the carrier frequency specified in the user prompt,  it calculates path loss for all possible links in the network. 
The analysis leverages \FuncSty{pycraf}'s implementation of International Telecommunication Union  Sector Radiocommunication (ITU-R)  propagation models combined with NASA SRTM terrain elevation data to account for realistic geographic features including topographic effects,  diffraction,  and atmospheric effects. 
For each potential link between nodes, the tool computes path loss values and applies Shannon's theorem to estimate achievable data rates with specified channel bandwidth. 
The output includes comprehensive link databases with path loss measurements,  capacity estimates,  and processed datasets formatted for the optimization stage.
The \FuncSty{GeographicDataCollectionTool} and \FuncSty{NetworkAnalysisTool} integrate diverse data modalities, including geospatial vector data from OpenStreetMap, population density raster maps from Meta, and terrain elevation data from NASA.

\subsubsection{\FuncSty{NetworkOptimizationTool}}
This tool transforms the network analysis results into actionable deployment decisions by solving a constrained optimization problem. 
Given the path loss information along with the cost constraints and performance objectives specified in the user prompt, it formulates and solves a mixed-integer programming problem using Google's OR-Tools library. 
The optimization jointly determines which base stations to deploy and how to route traffic through the network to minimize deployment cost while ensuring all demand nodes receive their required data rates. 
The tool handles complex constraints including budget limitations, minimum service requirements, and capacity bounds,  while balancing potentially conflicting objectives through weighted optimization. 
The output includes the final deployment plan specifying exact HAP and TBS locations,  active network links,  total deployment cost,  achieved data rates per node,  and 3D visualizations of the optimized network topology.

\subsection{Implementation and Open Source Availability}
To facilitate reproducibility and community adoption, we have open-sourced the complete MAINTAINED framework implementation on GitHub \cite{yzhang2025code}. 
The implementation utilizes Ollama for local deployment of the Alibaba's Qwen3-4B language model,  ensuring data privacy by eliminating dependencies on external LLM inference APIs.
The core agent implementation builds upon HuggingFace's \FuncSty{smolagents} project,  which provides robust infrastructure for function calling-based AI agents with standardized tool integration and ReAct reasoning capabilities.
Each network planning tool follows a modular architecture with formally defined input/output schemas that ensure seamless tool composition. 
This design enables domain experts to extend or replace individual tools without modifying the core agent logic, while maintaining full computational transparency.

\section{Case Study: Network Deployment in Saudi Arabia}\label{sec:ca_stu}
In this section, we demonstrate the MAINTAINED framework in action by applying it to  rural network deployment planning scenario in Saudi Arabia, and discuss its performance.
This planning task uses periodically updated datasets provided by OpenStreetMap, NASA SRTM, and Meta, as shown in  \figref{fig:framework}, which are appropriate for relatively stable variables such as topography and population distribution.

 \begin{figure}[!htp]
	\centering
	\includegraphics[width=\columnwidth]{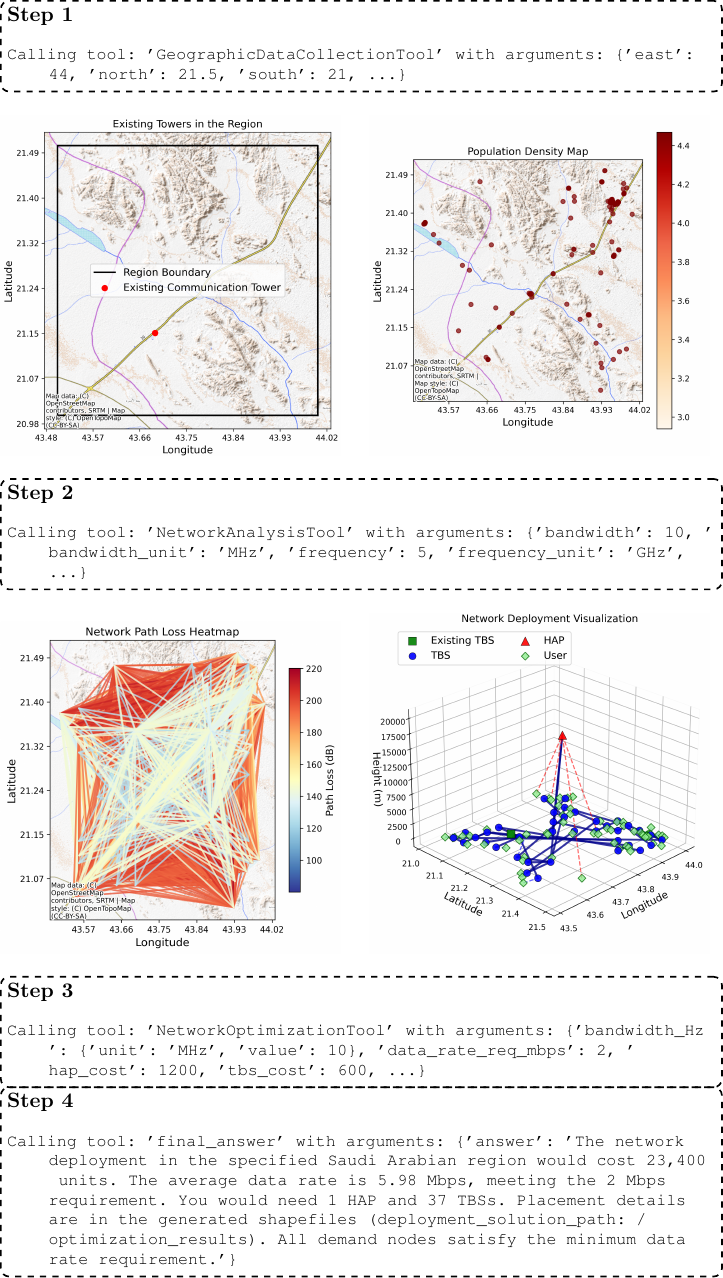}
	\caption{Execution workflow and output visualization of the proposed framework. }
	\label{fig:output}
\end{figure}

\subsection{Illustrative Workflow of the MAINTAINED Framework}
\figref{fig:output} illustrates the step-by-step workflow of our proposed framework for processing the example task introduced in  Sec.~\ref{sec:task}. 
A recording of this demonstration is available in \cite{yzhang2025demo}. 
As shown in the figure, the agent begins by retrieving geographic data through the \FuncSty{GeographicDataCollectionTool}, collecting essential information such as terrain maps, population density, and existing infrastructure. 
This establishes the spatial context necessary for network planning. Next, the agent employs the  \FuncSty{NetworkAnalysisTool} to analyze signal propagation at 5 GHz with 10 MHz bandwidth,  computing path loss and assessing candidate deployment sites under geographic and terrain constraints. 
Based on this analysis, the agent invokes the \FuncSty{NetworkOptimizationTool} to generate an optimal deployment plan that minimizes costs while guaranteeing a minimum of 2~Mbps coverage for all demand nodes.  
The solution includes one HAP and 37 TBSs, achieving full coverage at a total cost of 23,400 units. The agent then presents the results in a clear and concise manner, reporting that the deployment meets all service requirements with an average data rate of 5.98 Mbps. 
This example demonstrates how MAINTAINED integrates specialized tools into an autonomous workflow that efficiently delivers practical network deployment solutions.
 
\begin{table*}[!htp]
	\centering
	\caption{Performance comparison of wireless network planning approaches.}
	    \label{tab:performance}
	    \footnotesize
	    \setlength{\tabcolsep}{3pt}
	    \renewcommand{\arraystretch}{0.9}
	\begin{tabular}{|l|c|c|c|c|c|c|c|c|c|}
		\hline
		\textbf{Method} & \makecell{\textbf{TBS}\\\textbf{Count}} 
		& \makecell{\textbf{HAP}\\\textbf{Count}} 
		& \makecell{\textbf{Claimed}\\\textbf{Average Rate}\\\textbf{(Mbps)}}
		& \makecell{\textbf{Deployment}\\\textbf{Cost (units)}} 
		& \makecell{\textbf{Verified}\\\textbf{Average Rate}\\\textbf{(Mbps)}}
		& \makecell{\textbf{Verified Efficiency}\\\textbf{(bps/unit)}}
		& \textbf{Availability}
		& \makecell{\textbf{GPU}\\\textbf{Memory}} 
		& \makecell{\textbf{Inference}\\\textbf{Cost Level}} \\
		\hline
		ChatGPT-4o & 50 & 5 & 48.2 & 36,000 & 0.42 & 11.7 & API Only & -  & High \\
		\hline
		Claude Sonnet 4 & 2 & 1 & 9.0 & 2,400 & 0.08 & 33.3 & API Only & - & High \\
		\hline
		DeepSeek-R1 & 0 & 2 & 2.78 & 1,200 & 0.19 & 158.3 & Open Source & ~140GB & Medium \\
		\hline
		Qwen3-4B & 200  & 1  & 10.0 & 121,200 & 4.5 & 38.1 & Open Source & 8GB & Low \\
		\hline
		RAG-Enhanced & 30 & 15 & 3.5 & 36,000 & 3.0 & 83.3 & Open Source & 12GB & Low \\
		\hline 
		\rowcolor[HTML]{F2F4FF}
		\textbf{MAINTAINED} & \textbf{37} & \textbf{1} & \textbf{5.98} & \textbf{23,400} & \textbf{5.98} & \textbf{255.6} & \textbf{Open Source} & \textbf{8GB} & \textbf{Low} \\
		\hline
	\end{tabular}
\end{table*}

\subsection{Performance Comparison and Key Takeaways}
\label{sec:perf_comp}

Since classical methods require manual expertise, MAINTAINED prioritizes the autonomous orchestration of reliable algorithms.
Consequently, we evaluate performance against state-of-the-art LLMs and RAG baselines rather than classical methods.
The results, summarized in Table~\ref{tab:performance}, reveal notable disparities in performance across these approaches. 
Pure LLM-based methods exhibit significant discrepancies between their claimed and verified performance. 
For instance,  ChatGPT-4o,  despite deploying 50 TBSs and 5 HAPs at a cost of 36,000 units,  achieves a verified average data rate of only 0.42 Mbps,  which is far below its claimed 48.2 Mbps. 
Claude Sonnet 4 and DeepSeek-R1 show similar shortfalls, with verified rates of just 0.08 Mbps and 0.19 Mbps,  respectively. 
These results highlight a key limitation of purely parametric LLMs: they often rely on oversimplified assumptions,  such as uniform grid-based user distribution,  and fail to account for realistic geographic and terrain variations. 
Although Qwen3-4B achieves a more accurate verified rate of 4.5 Mbps,  it incurs an impractically high deployment cost of 121,200 units due to excessive overprovisioning of 200 TBSs.

The RAG-Enhanced baseline, which is built on top of Qwen3-4B with additional retrieval mechanisms, achieves a more balanced outcome by narrowing the gap between predicted and actual performance. 
It delivers a verified rate of 3.0 Mbps at a deployment cost of 36,000 units. 
While it is capable of retrieving relevant contextual information,  it cannot accurately model terrain-aware path loss. 
In contrast,  our proposed MAINTAINED framework achieves the best overall results,  with a verified average data rate of 5.98 Mbps and the highest efficiency at 255.6 bps per unit cost.
Among all the frameworks where the verified average data rate meets the requirements, our proposed MAINTAINED framework achieves at least a 35\% reduction in deployment cost and delivers the highest verified average data rate.
Moreover, MAINTAINED is highly resource-efficient. 
It requires only 8 GB of graphics processing unit (GPU)  memory, significantly less than the 140 GB required by DeepSeek-R1, and is therefore well-suited for edge deployment.
These results underscore a fundamental insight: effective wireless network planning requires seamless integration of LLM reasoning capabilities with domain-specific computational tools.
While pure LLMs excel at high-level planning and user interaction, they cannot replace physics-based modeling and optimization algorithms essential for accurate network design. 
Although performing these essential computations increases our agent’s end-to-end execution time to 159 seconds (see \cite{yzhang2025demo}), it yields a verified and accurate solution. By contrast, traditional LLMs achieve low latency precisely by omitting these computations, often producing the invalid results reported in Table \ref{tab:performance}.

\section{Conclusion and Future Directions}\label{sec:concl}
This work introduces a paradigm shift in applying AI to wireless communications by demonstrating that verified computation through tool augmentation enables compact LLMs to outperform much larger ones in technical tasks. 
The proposed MAINTAINED framework shows that a 4B-parameter model, when equipped with specialized tools,  can deliver superior wireless network planning performance compared to models with hundreds of billions of parameters. 
Beyond wireless communications,  this work suggests a broader shift in how AI should be applied to engineering domains. Rather than scaling up models endlessly,  future progress lies in coupling reasoning agents with reliable external computation. 
This hybrid strategy enhances transparency, reduces inference cost, and improves accuracy in practical deployments.

Building on this foundation, several promising research directions emerge: 
1) assessing the agent's generalization and reasoning capabilities by testing its orchestration of held-out tools, such as using deterministic ray-tracing;
2) the tool-augmented paradigm can be extended to other 6G functions, such as dynamic spectrum access,  network slicing,  and low-latency scheduling.

\bibliographystyle{IEEEtran}
\bibliography{IEEEabrv,hokie-HD}

@Software{winkel_pycraf_2018,
  author       = {Winkel, Benjamin and others},
  howpublished = {\url{https://github.com/bwinkel/pycraf}},
  title        = {pycraf: A package that provides functions and procedures for various tasks in spectrum-management compatibility studies},
  version      = {v2.1.0},
}

@Misc{smolagents,
  author       = {Aymeric Roucher and others},
  howpublished = {\url{https://github.com/huggingface/smolagents}},
  title        = {smolagents: a smol library to build great agentic systems.},
  year         = {2025},
}

@Article{Zhou2025,
  author    = {Zhou, Hao and others},
  journal   = {IEEE Wirel. Commun.},
  title     = {Large Language Models for Wireless Networks: An Overview from the Prompt Engineering Perspective},
  year      = {2025},
  issn      = {1558-0687},
  pages     = {1--9},
  doi       = {10.1109/mwc.001.2400384},
  fjournal  = {IEEE Wireless Communications},
  publisher = {Institute of Electrical and Electronics Engineers (IEEE)},
}

@Article{Ji2023,
  author    = {Ji, Ziwei and others},
  journal   = {ACM Comput. Surv.},
  title     = {Survey of Hallucination in Natural Language Generation},
  year      = {2023},
  issn      = {1557-7341},
  month     = dec,
  number    = {12},
  pages     = {1--38},
  volume    = {55},
  fjournal  = {ACM Computing Surveys},
  publisher = {ACM},
}

@InProceedings{Yao2023,
  author    = {Yao, Shunyu and others},
  booktitle = {Proc. ICLR},
  title     = {{ReAct}: Synergizing Reasoning and Acting in Language Models},
  year      = {2023},
  month     = may,
  doi       = {10.48550/ARXIV.2210.03629},
}

@Article{Bariah2024a,
  author    = {Bariah, Lina and others},
  journal   = {IEEE Wirel. Commun.},
  title     = {{AI} Embodiment Through {6G}: Shaping the Future of {AGI}},
  year      = {2024},
  issn      = {1558-0687},
  month     = oct,
  number    = {5},
  pages     = {174--181},
  volume    = {31},
  doi       = {10.1109/mwc.015.2300521},
  fjournal  = {IEEE Wireless Communications},
  publisher = {Institute of Electrical and Electronics Engineers (IEEE)},
}

@Misc{meta_hrsl_2023,
  author       = {{Meta}},
  howpublished = {\url{https://dataforgood.facebook.com/dfg/tools/high-resolution-population-density-maps}},
  title        = {High Resolution Population Density Maps},
}

@Article{Boeing2025,
  author    = {Boeing, Geoff},
  journal   = {Geogr. Anal.},
  title     = {Modeling and Analyzing Urban Networks and Amenities With {OSMnx}},
  year      = {2025},
  issn      = {1538-4632},
  month     = may,
  doi       = {10.1111/gean.70009},
  fjournal  = {Geographical Analysis},
  publisher = {Wiley},
}

@Misc{yzhang2025demo,
  author       = {Zhang, Yongqiang},
  howpublished = {\url{https://www.youtube.com/watch?v=DKK1pIPGMQc}},
  title        = {Code Execution and Output Demo},
}

@Misc{yzhang2025code,
  author       = {Zhang, Yongqiang},
  howpublished = {\url{https://github.com/antsfromupthere/maintained}},
  title        = {{MAINTAINED}: Tool-Augmented {AI} for Wireless Network Deployment},
}

@InProceedings{Zhang2024c,
  author    = {Zhang, Han and others},
  booktitle = {Proc. IEEE GLOBECOM},
  title     = {Large Language Models in Wireless Application Design: In-Context Learning-enhanced Automatic Network Intrusion Detection},
  year      = {2024},
  month     = dec,
  pages     = {2479--2484},
  doi       = {10.1109/globecom52923.2024.10901312},
}

@Article{Hou2025,
  author   = {Hou, Jinbo and others},
  journal  = {IEEE Commun. Mag.},
  title    = {{iPLAN}: Redefining Indoor Wireless Network Planning Through Large Language Models},
  year     = {2025},
  note     = {to appear},
  fjournal = {IEEE Communications Magazine},
}

@Article{Jiang2025,
  author    = {Jiang, Feibo and others},
  title     = {From Large {AI} Models to Agentic {AI}: A Tutorial on Future Intelligent Communications},
  year      = {2025},
  month     = may,
  note      = {available online: https://arxiv.org/abs/2505.22311},
  publisher = {arXiv},
}

@Article{Quan2025,
  author    = {Quan, Hongye and others},
  journal   = {IEEE Netw. Lett.},
  title     = {Large Language Model Agents for Radio Map Generation and Wireless Network Planning},
  year      = {2025},
  issn      = {2576-3156},
  month     = sep,
  number    = {3},
  pages     = {166--170},
  volume    = {7},
  doi       = {10.1109/lnet.2025.3539829},
  publisher = {Institute of Electrical and Electronics Engineers (IEEE)},
}

@Article{Tong2025,
  author    = {Tong, Jingwen and others},
  title     = {{WirelessAgent}: Large Language Model Agents for Intelligent Wireless Networks},
  year      = {2025},
  month     = may,
  note      = {available online: https://arxiv.org/abs/2505.22311},
  publisher = {arXiv},
}

@Article{Zou2025,
  author    = {Zou, Hang and others},
  journal   = {IEEE Trans. Mach. Learn. Commun. Netw.},
  title     = {{TelecomGPT}: A Framework to Build Telecom-Specific Large Language Models},
  year      = {2025},
  issn      = {2831-316X},
  pages     = {948--975},
  volume    = {3},
  doi       = {10.1109/tmlcn.2025.3593184},
  fjournal  = {IEEE Transactions on Machine Learning in Communications and Networking},
  publisher = {Institute of Electrical and Electronics Engineers (IEEE)},
}
\vspace{-1.0cm}
\begin{IEEEbiographynophoto}{Yongqiang Zhang} 
	received the Ph.D. degree from King Abdullah University of Science and Technology (KAUST), Saudi Arabia. 
\end{IEEEbiographynophoto} 
\vspace{-1.0cm}
\begin{IEEEbiographynophoto}{Mustafa A. Kishk} is an assistant professor at Maynooth University, Ireland.
\end{IEEEbiographynophoto}
\vspace{-0.9cm}
\begin{IEEEbiographynophoto}{Mohamed-Slim Alouini} 
	 is a distinguished professor at  KAUST.
\end{IEEEbiographynophoto} 
\end{document}